\documentclass[preprint,12pt]{elsarticle}

\usepackage{amssymb}
\makeatletter
\usepackage{mathrsfs}
\usepackage{multirow}
\usepackage{tabularx}
\usepackage{threeparttable}
\usepackage{amsmath,bm}
\usepackage{amsmath}
\usepackage{graphicx}
\usepackage{graphics}
\usepackage{epstopdf}
\newtheorem{theorem}{Theorem}
\newtheorem{remark}{Remark}
\newtheorem{lemma}{Lemma}
\newtheorem{corollary}{Corollary}
\usepackage[justification=centering]{caption}

\newcommand{\Rmnum}[1]{\expandafter\@slowromancap\romannumeral #1@}
\makeatother
\begin{document}
\begin{frontmatter}
\title{Threshold Changeable Secret Sharing Scheme and Its Application to
Group Authentication}
\author{Keju Meng$^*$}
\author{Yue Yu}
\author{Fuyou Miao}
\author{Wenchao Huang}
\author{Yan Xiong}
\address{University of Science and Technology of China, No.96, JinZhai Road Baohe District,Hefei,Anhui, 230026,P.R.China}
\cortext[cor]{Corresponding author.}
\begin{abstract}
Group oriented applications are getting more and more popular in mobile Internet and call for secure and efficient secret sharing (SS) scheme to meet their requirements. A $(t,n)$ threshold SS scheme divides a secret into $n$ shares such that any $t$ or more than $t$ shares can recover the secret while less than $t$ shares cannot. However, an adversary, even without a valid share, may obtain the secret by impersonating a shareholder to recover the secret with $t$ or more legal shareholders. Therefore, this paper uses linear code to propose a threshold changeable secret sharing (TCSS) scheme, in which threshold should increase from $t$ to the exact number of all participants during secret reconstruction. The scheme does not depend on any computational assumption and realizes asymptotically perfect security. Furthermore, based on the proposed TCSS scheme, a group authentication scheme is constructed, which allows a group user to authenticate whether all users are legal group members at once and thus provides efficient and flexible m-to-m authentication for group oriented applications.
\end{abstract}
\begin{keyword} Threshold Changeable Secret Sharing, Asymptotically Perfect \sep 
Linear Code, Illegal Participant Attack, Group Authentication 
\end{keyword}
\end{frontmatter}

\section{Introduction}
With the development of mobile Internet, network applications do not limit to 1-to-1 or 1-to-m (i.e., client/server) interaction pattern any more. Group oriented applications with m-to-m interaction pattern are getting more and more popular especially in mobile social apps. Group chat is one of group oriented applications provided by most social apps, and it is actually an online meeting system and allows a user to invite his/her friends to have a meeting anytime and anywhere. For example, WeChat, the most popular mobile social app with over 600 million users in Asia, enables a user to initiate a group for an ad hoc session on demand. In the group, any user is allowed to send/receive messages, start an audio/video chat or invite his/her friends into the group. Consequently, a user may not be familiar with some other users. However, a main concern about the online meeting is authentication. That is, each user needs to make sure that any other user in the group has the right identity, especially when the meeting is confidential. That is because any user at a confidential meeting is responsible for the information he/she releases, and never wants any wrong person to have access to it. In this case, each user needs to authenticate all the others successfully. Otherwise, the meeting must be aborted. Traditionally, one user employs a 1-to-1 authentication scheme to verify another user's identity. In such a scheme, one user (verifier) gets convinced that the other user (prover) is the right one it claims to be. If the 1-to-1 scheme is trivially applied to mutual authentication within a group of $m$ users, there are totally $m(m-1)$ rounds of authentication. Nevertheless, $m$ rounds are sufficient for the same case if m-to-m authentication scheme is employed, because the new authentication allows each user to verify whether all users are legal group members at once. Therefore, it is of great importance to find a proper cryptographic tool in designing secure and efficient m-to-m authentication schemes for group oriented applications.\par
As a group oriented cryptographic primitive, $(t,n)$ threshold secret sharing (SS) scheme  divides a secret into $n$ shares and allocates each share to a shareholder, such that at least $t$ shareholders are qualified to recover the secret but less than $t$ are not. In a group of exact $t$ shareholders, each shareholder can verify whether all $t$ shareholders are legal at once if they mutually exchange shares, independently reconstruct the secret and check the correctness. Therefore, $(t,n)$ threshold SS has potential in building m-to-m authentication schemes. \par
Since $(t,n)$ threshold SS was first introduced by Shamir \cite{shamir1979share} and Blakley \cite{blakley1979safeguarding} in 1979, it has been studied extensively and widely used in many applications, such as secure multiparty computation \cite{boyle2018foundations}, threshold signature \cite{lu2017forward}, group key agreement \cite{meng2019secure}, group authentication \cite{harn2012group} and so on. As the most popular $(t,n)$ threshold SS, Shamir's scheme \cite{shamir1979share} is constructed based on a polynomial of degree at most $(t-1)$ in a finite field. Blakley's scheme \cite{blakley1979safeguarding} is based on hyperplane while Asmuth-Bloom's scheme \cite{asmuth1983modular} and Mignotte's scheme \cite{mignotte1982share} are both based on Chinese remainder theorem. Linear code is another tool to construct $(t,n)$ SS schemes. In 1981, McEliece and Sarwate \cite{mceliece1981sharing} proposed a formulation of $(t,n)$ threshold secret sharing scheme based on maximum-distance-separable (MDS) codes, and pointed out that Shamir's $(t,n)$ threshold SS scheme can be constructed equivalently by using Reed-Solomon code. Subsequently, many secret sharing schemes based on linear codes were proposed \cite{carlet2005linear,massey1993minimal,massey1995some}. In \cite{massey1993minimal} and \cite{massey1995some}, Massey utilized the linear code to construct a secret sharing scheme. Meanwhile, he also presented the relationship between the access structure of secret sharing and the minimal code word of the dual code in linear code. All the above $(t,n)$ threshold SS schemes do not depend on any computational assumption of hard problem or one way function. \par
\begin{figure}
  \centering
  \includegraphics[scale=0.3]{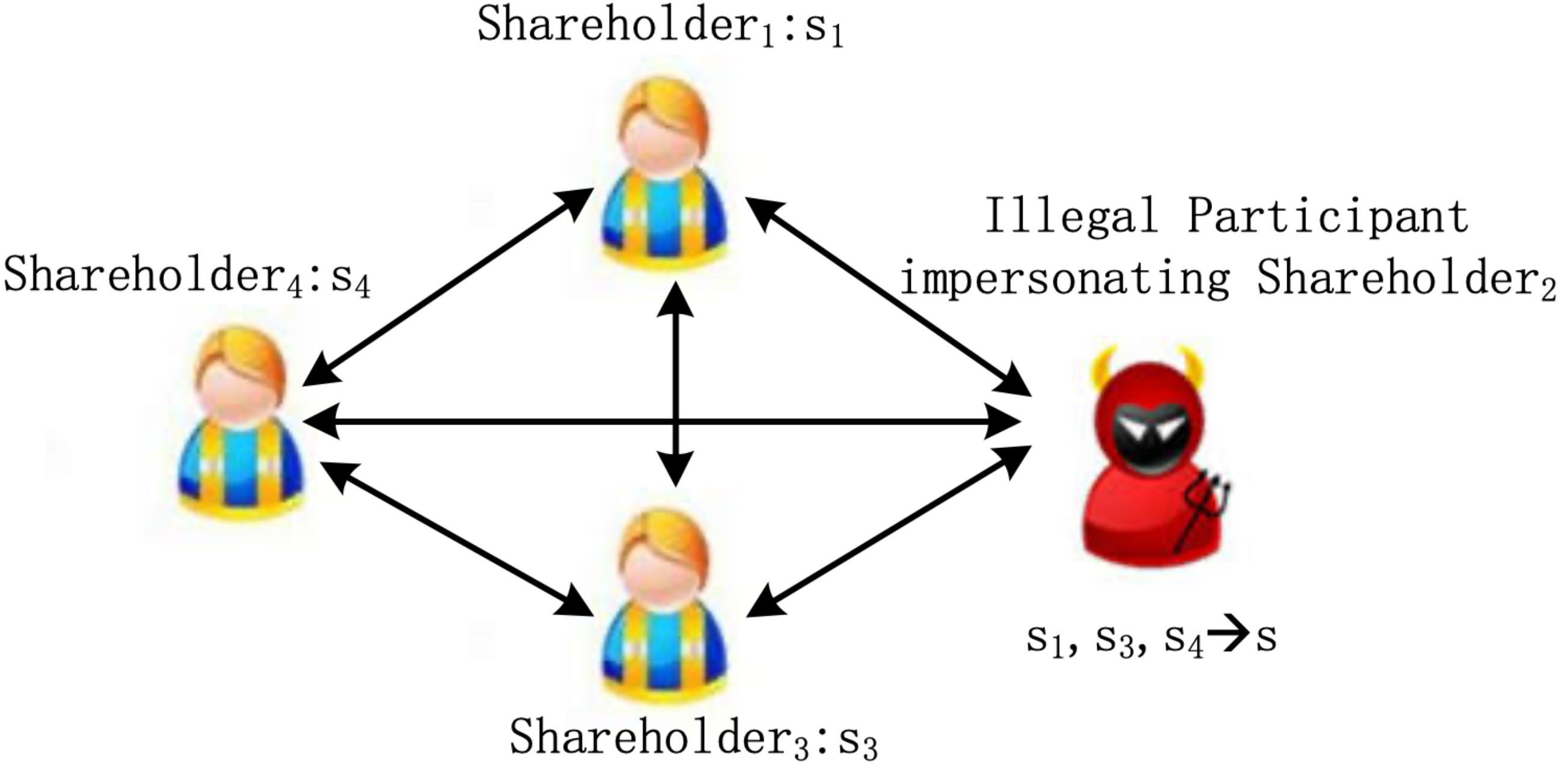}
   \caption{Example of IP attack with threshold 3} \label{Fig1}
\end{figure}

In general, a $(t,n)$ threshold SS scheme consists of share generation and secret reconstruction. In share generation, the dealer generates $n$ shares from the secret to be shared and allocates each shareholder a share securely. In secret reconstruction, $t$ or more than $t$ shareholders exchange shares privately and thus each shareholder can recover the secret from collected shares. In a $(t,n)$ threshold SS scheme, if exact $t$ shareholders pool shares together to recover the secret, all the shares must be valid. However, when the number of participants in secret reconstruction is more than $t$, Tompa and Woll \cite{tompa1989share} proposed an attack model to traditional (t,n) threshold SS schemes. Suppose there are $m$ ($m > t$) participants in secret reconstruction, and one of them is an adversary without any valid share, it may lead some potential danger. The illegal participant in the name of a legal shareholder may communicate with the others. As a result, it could collect enough ($m-1 \ge t$) valid shares from the other participants and recover the secret. The attack model is called illegal participant attack (IPA) and Fig \ref{Fig1} is an example of IPA with $t=3$.

One countermeasure against IPA is cheater detection and the typical scheme is verifiable secret sharing (VSS) \cite{peng2016publicly,rajabi2019verifiable}. In a VSS scheme, when a participant receives shares from others in secret reconstruction, it must verify validity of the shares at first. Obviously, the illegal participant without valid share can be distinguished from all the participants. Although VSS or some other cheater detection schemes \cite{banerjee2018hierarchy,liu2016linear,xu2015note} can thwart IPA to some degree, the illegal shareholder still can steal secret because when the adversary sends invalid shares to the others, legal shareholders also send valid shares to the adversary. \par
The other countermeasure is cheater immune \cite{martin2008challenging} and the typical scheme is threshold changeable (also known as dynamic threshold) secret sharing (TCSS). TCSS allows shareholders to change threshold from $t$ to $t'$ where $t' > t$. In this case, $t'$ or more than $t'$ shareholders can recover the secret while less than $t'$ cannot. Therefore, some TCSS schemes can prevent IPA if there are exact $t'$ participants with $t\le t'\le n$ and the threshold is also raised to $t'$. For example, the dealers in schemes \cite{martin1999changing}, generate multiple shares for every shareholder, each share with a distinct threshold $t'>t$. Therefore, if shares with $t'$ threshold are used to reconstruct the secret, all the $t'$ shares must be valid. Otherwise, the secret cannot be recovered, which means the scheme can prevent IPA. However, every shareholder in such schemes has to hold multiple shares and thus requires more storage. Moreover,the scheme only raises threshold to predefined values. As an improvement, Ron Steinfeld et al. presented two lattice-based TCSS schemes \cite{steinfeld2006lattice,steinfeld2007lattice}, in which shareholders add some noise to their shares or delete some bits of their shares to compute subshares which contain partial information about the original shares. As a result, a larger number $t'>t$ of subshares are required to recover the secret by solving the closest vector problem (CVP). The two schemes do not require communication either between the dealer and shareholders or among shareholders. However, 1) the share combiner needs to communicate with all participants and instruct them to change threshold; 2) the scheme is lattice-based and thus is complicated in computation; 3) it is far from a perfect scheme. In 2017, H. Pilaram and T. Eghlidos proposed a lattice based threshold changeable multi-secret sharing scheme \cite{pilaram2017lattice}. Nojoumian et al. \cite{nojoumian2009dealer} presented a Shamir's $(t,n)$ SS based dealer-free TCSS scheme using secure multiparty computation. The scheme remains some properties of Shamir's $(t,n)$ threshold SS, such as being unconditionally secure and ideal. Moreover, it can change the threshold to any value. Later on, they also proposed a method \cite{nojoumian2013dealer} of increasing threshold by zero addition. The scheme increases the threshold by generating shares of a polynomial that corresponds to a secret with value zero and threshold $t'>t$, and adding these new shares to player's current shares.  However, due to resharing operation, both schemes require too much computation in participants and communication among participants. In 2016, Yuan et al. \cite{Yuan2016Novel} came up with two TCSS schemes based on Lagrange interpolation polynomial and 2-variable one-way functions. Both schemes require the dealer to evaluate and store a lot of values before increasing the threshold. Moreover, it needs the combiner (i.e., the proxy of the dealer) to send each participant a distinct key to active the share additionally. Harn and Hsu utilized linear combination of multi univariate polynomials or   bivariate polynomial to implement two TCSS schemes \cite{harn2012group,harn2015dynamic}. However, Ahmadian and Jamshidpour's works \cite{ahmadian2017linear,jamshidpour2017security} has employed linear subspace method to attack \cite{harn2012group,harn2015dynamic} successfully. Besides, Jia et al. proposed a TCSS scheme \cite{jia2019new} based on Chinese remainder theorem. Because the scheme incorporates Krawczyk’s encryption \cite{krawczyk1993secret}, it is just computational security.\par
Based on the above, we summarize our contributions in two aspects.
\begin{enumerate}
\item This paper proposes a TCSS scheme based on linear code. The scheme does not depend on any hard problem or one-way function and realizes asymptotically perfect.
\item As an application of the TCSS scheme, a group authentication protocol is constructed to enable the rapid m-to-m authentication in group oriented applications.
\end{enumerate}

The rest of this paper is organized as follows. In next section, we briefly review some notations, definitions and linear code based $(t,n)$ threshold SS scheme. We describe our proposed TCSS scheme in Section 3. Section 4 analyzes the security and summarizes the properties of the TCSS scheme. As an application, a group authentication protocol is constructed in Section 5. Finally, we conclude the paper in Section 6.
\section{Preliminaries} 
\subsection{Notations and terms}
\textbf{1) Notations}\par
Here are some notations used throughout the paper. $ {{I_n}} $ denotes the integer set $\{ 1,2,...,n\} $ and is used to label all the $n$ shareholders. $ {{I_m}} $, with the cardinality $|{I_m}| = m$, $(t \le m \le n)$, is a subset of $ {{I_n}} $. $ {{I_m}} $ is used to label any $m$ out of $n$ shareholders. ${F_p} = \{ 0,1,2,...,p - 1\} $ is a finite field for large prime $p$. $F_p^* = \{ 1,2,...,p - 1\} $ is the multiplicative group of ${F_p}$. $r{ \in _U}{F_p}$ denotes that $r$ is a random number uniformly distributed in ${F_p}$. $|S|$ is the cardinality of set $S$. \par
\textbf{2) Information theoretical terms} \par
Now, we introduce some basic terms in information theory. Suppose $X,Y$ are discrete-time discrete-valued random variables with sample space ${\rm{S}}{{\rm{P}}_{\rm{1}}},{\rm{S}}{{\rm{P}}_{\rm{2}}}$. The entropy of $X$ is denoted as \[H(X) = E( - {\log _2}P(X)) = \sum\limits_{x \in {\rm{S}}{{\rm{P}}_{\rm{1}}}} { - P(x){{\log }_2}P(x)}, \] where $E$ is the expectation operator and $P({\rm{X}})$ is the probability distribution function of $X$. From the view of an adversary, the secret $s$ in $(t,n)$ threshold SS is indistinguishable from a random variable uniformly distributed in secret space. Therefore, we use $H(s)$ to denote uncertainty of the secret. \par
The mutual information of $X$ with $Y$ is written as 
\begin{align*}
I(X;Y) & = H(X) - H(X/Y) \\
& = \sum\limits_{x \in {\rm{S}}{{\rm{P}}_{\rm{1}}},y \in {\rm{S}}{{\rm{P}}_{\rm{2}}}} {P(xy){{\log }_2}P(x/y)} /P(x).
\end{align*}
$I(X;Y)$ means the amount of information about $X$ obtained due to the knowledge $Y$. In the following sections, we will write ${\log _2}P(x)$ as $\log P(x)$ for simplicity. 
\subsection{$(t,n)$ threshold SS based on linear code}
There are several ways to construct a $(t,n)$ threshold SS scheme based on linear code \cite{massey1993minimal}, and one of them comes as follows: \par
Assume that a $[n + 1,t]$ linear code $\mathbf{LC}$ is a subspace of $F_p^{n + 1}$ with length $n+1$ and dimension $t$, and $G = ({\vec g_0},{\vec g_1},...{\vec g_n})$ is the public generator matrix of linear code $\mathbf{LC}$, where ${\vec g_i} \in F_p^t$ $(0 \le i \le n)$ is a nonzero column vector. $G$ has the rank $t$. In the traditional $(t,n)$ threshold SS scheme based on $\mathbf{LC}$, there is a dealer and $n$ shareholders ${U_1},{U_2},...,{U_n}$ and the secret $s$ is a value in ${F_p}$. The scheme consists of the following two steps: \par
\textbf{Share Generation:} The dealer privately chooses a row vector $\vec v = ({v_0},{v_1},...,{v_{t - 1}})  \in F_p^t$, such that the secret is $s = \vec v{\vec g_0}\bmod p$. It is obvious that there are totally ${p^{t - 1}}$ such $\vec v$ for a given pair $(s,{\vec g_0})$. The dealer generates the code word $\vec w = ({s_0},{s_1},...,{s_n}) = \vec vG\bmod p$ and allocates ${s_i} = \vec v{\vec g_i}\bmod p$ to ${U_i}$ as the share securely, for $i = 1,2,...,n$. \par
\textbf{Secret Reconstruction:} If $m$, ($m \ge t$) out of $n$ shareholders, ${\mathbf{\mathcal{U}_{{I_m}}}} = \{ {U_{{i_1}}},{U_{{i_2}}},...,{U_{{i_m}}}\} $ need to recover the secret $s$, they are supposed to find a group of parameters $\{ {b_{{i_1}}},{b_{{i_2}}},...,{b_{{i_m}}}\} $ over ${F_p}$ such that ${\vec g_0} = {b_{{i_1}}}{\vec g_{{i_1}}} + {b_{{i_2}}}{\vec g_{{i_2}}} + ... + {b_{{i_m}}}{\vec g_{{i_m}}}\bmod p$ holds, and then pool their shares $\{ {s_{{i_1}}},{s_{{i_2}}},...,{s_{{i_m}}}\} $ in private to compute the secret as \[s = \vec v{\vec g_0} = \sum\limits_{j = 1}^m {{b_{{i_j}}}\vec v{{\vec g}_{{i_j}}}}  = \sum\limits_{j = 1}^m {{b_{{i_j}}}{s_{{i_j}}}} \bmod p.\]
\subsection{Definitions}
\textbf{Definition 2.1.} (Perfect $(t,n)$ threshold SS) Let $s$, $S$ and $\Omega $ be the secret, secret space and the share set of a $(t,n)$ threshold SS with $|\Omega | = n$. The $(t,n)$ threshold SS is perfect with respect to probability distribution of $s$ on the secret space $S$ if \par
1) $H(s) \ge 0$, \par
2) $I(s;{\Omega _J}) = H(s) - H(s|{\Omega _J}) = 0$, \\
where ${\Omega _J}$ denotes any subset of $\Omega $ with less than $t$ shares, i.e. ${\Omega _J} \subseteq \Omega $ and $|{\Omega _J}| < t$. \par
As a secret value, the secret $s$ actually appears as a random variable uniformly distributed in secret space $S$. In a perfect $(t,n)$ threshold SS, less than $t$ shareholders get no information about the secret even if they have up to $t-1$ shares. Loosening the perfect $(t,n)$ threshold SS a little bit, we get the definition of asymptotically perfect $(t,n)$ threshold SS as follows. \par
\textbf{Definition 2.2.} (Asymptotically perfect $(t,n)$ threshold SS) A $(t,n)$ threshold SS is asymptotically perfect with respect to probability distribution of $s$ on secret space $S$ if, for all ${\Omega _J}$ with $|{\Omega _J}| < t$, we have \par
1) $H(s) \ge 0$ \par
2) $\lim_{|S|\rightarrow\infty} I(s;{\Omega _J}) = 0$\\
where $|S|$ is the cardinality of $S$.\par
Asymptotically perfect $(t,n)$ threshold SS implies that less than $t$ shareholders get nearly no information about the secret when the secret space converges to infinity.\par 
\textbf{Definition 2.3} (Threshold changeable SS) Formally, let $S$, ${S_H}$ and ${I_D}$ be the secret space,  share space and identity space respectively in $(t,n)$ threshold SS. $s \in S$ is the secret. Assume ${\mathcal{U}} = \{ {U_i}|{U_i} \in {I_D},i = 1,2,...,n\} $ are $n$ shareholders. Each shareholder ${U_i}$ has the private share ${s_i} \in S_H$ and public identity ${U_i}$. \par
Before any $m$ shareholders ${{\mathcal{U}}_m}$ (i.e., ${{\mathcal{U}}_m} \subseteq {\mathcal{U}}$ and ${\rm{|}}{{\mathcal{U}}_m}| = m$) want to recover the secret, they need to raise the threshold from $t$ to $m$ by constructing a component. That is, each shareholder ${U_j} \in {{\mathcal{U}}_m}$ constructs a component ${c_j} = f({s_j},{r_j},{{\mathcal{U}}_m})$, where $f:{S_H} \times S \times SUB({I_D}) \rightarrow {S_H}$ is a component construction function. ${s_j}$ is the original share of ${U_j}$ and $c_j$ is the component matched with threshold $m$. ${r_j}$ is random number uniformly selected from $S$, i.e., ${r_j}{ \in _U}S$, and $SUB({I_D})$ denotes the power set of ${I_D}$. Therefore, ${\mathcal{C}_m} = \{ {c_j} = f({s_{{j_{}}}},{r_j},{{\mathcal{U}}_m})|{U_j} \in {{\mathcal{U}}_m}\} $ is a valid component set of ${{\mathcal{U}}_m}$. The $(t,n)$ threshold SS is TCSS if 
\begin{equation} \label{Equation1}
I(s;{\mathcal{C}}') = \left\{ \begin{array}{l}
H(s)\mathop {}\nolimits^{^{\mathop {}\nolimits^{} }} \mathop {}\nolimits^{} i{f^{}}{\mathcal{C}}' = {{\mathcal{C}}_m},t \le |{{\mathcal{C}}_m}| = m \le n;\\
0 \  or \Rightarrow0{\mathop {}\nolimits^{} _{}}\mathop {}\nolimits_{} i{f^{}}{\mathcal{|C}}' \cap {{\mathcal{C}}_m}| < m.
\end{array} \right.
\end{equation}
where the secret $s$ is viewed as a random variable in $S$. ${{\mathcal{C}}}'$ is a component set actually used in recovering $s$, and $\Rightarrow0$ denotes converging to 0 if $|S|$ approaches to infinity. \par
Equation \ref{Equation1} implies the two facts.
\begin{enumerate}
\item If ${\mathcal{C}}'$, the component set actually used in secret reconstruction, is identical with ${{\mathcal{C}}_m}$ and the number of participants is no less than $t$, the secret is bound to be recovered. \
\item If ${\mathcal{C}}'$ does not contain all components in ${{\mathcal{C}}_m}$, almost no information about the secret can be obtained. In other words, the threshold is changed from $t$ to $m$.
\end{enumerate}
\section{Proposed TCSS Scheme Based on Linear Code}
\subsection{Entities and security goals} 
There are three types of entity in our proposed TCSS scheme, the dealer, $n$ shareholders and some adversaries.  \par
\emph{1)	Dealer:} The dealer is the honest coordinator trusted by all shareholders, and responsible for scheme setup such as determining system parameters, choosing the secret, generating and distributing shares and so on. \par
\emph{2)	Shareholders:} There are totally $n$ shareholders, each with a share generated by the dealer. When $m$ $(m \ge t)$ shareholders recover the secret, each of them generates a component matched with threshold $m$ and exchanges component with the others. Then, every shareholder independently recovers the secret from all components. \par
\emph{3) Adversary:} An adversary can impersonate a legal shareholder but without the right share. That is, in the name of some legal shareholder, it is allowed to receive private information (i.e., components in our scheme) from the other participants and send a forged component to the others. In this way, the illegal adversary obtains valid components and tries to use them to recover the secret. The IPA model is also mentioned in Section 1.\par
The TCSS scheme has two security goals.
\begin{enumerate}
\item Less than $t$ shareholders cannot recover the secret.\ 
\item The scheme can thwart IPA. Concretely, if an adversary uses a forged component to participate in secret reconstruction, anyone cannot recover the secret.
\end{enumerate}
\subsection{Our Scheme} \label{scheme}
Our TCSS scheme consists of three algorithms, share generation \textbf{SG}${(s,\mathcal{U})}$, component construction \textbf{CC}${\rm{(}}{{\mathcal{U}}_m},{\Omega _m},{R_m})$, and secret reconstruction \textbf{SR}${\rm{(}}{{\mathcal{C}}_m})$. \par
\textbf{SG}${(s,\mathcal{U})}$ takes the secret $s$ and the set of $n$ shareholders, $\mathcal{U}$, as input and generates $\Omega $, the set of $n$ shares as output. In this algorithm, the dealer generates $n$ shares from the secret $s$ and allocates each share to the corresponding shareholder, securely.\par
\textbf{CC}${\rm{(}}{{\mathcal{U}}_m},{\Omega _m},{\mathcal{R}_m})$ takes ${{\mathcal{U}}_m}$, ${\Omega _m}$ and ${{\mathcal{R}}_m}$ as input and outputs ${{\mathcal{C}}_m}$. ${{\mathcal{U}}_m}$ is the set of participants, denoting a subset of $\mathcal{U}$ with $m$ shareholders. ${\Omega _m}$ is the share set of ${{\mathcal{U}}_m}$. ${{\mathcal{R}}_m}$ is a set of $m$ random numbers and ${{\mathcal{C}}_m}$ is the set of $m$ components. In this algorithm, each participant in ${{\mathcal{U}}_m}$ generates a component with its share in ${\Omega _m}$ and a random number in ${{\mathcal{R}}_m}$ non-interactively.  \par
\textbf{SR}${\rm{(}}{{\mathcal{C}}_m})$ takes ${{\mathcal{C}}_m}$ as input and recovers the secret $s$ as output. In this algorithm, each participant uses all the $m$ components of ${{\mathcal{U}}_m}$ (i.e., ${\mathcal{C}}_m$) to recover the secret independently.\par
More detailed description are given as follows and the algorithm is also given in Figure \ref{Fig2}. \\
\textbf{1)	Share Generation} \par
Suppose there are $n$ shareholders ${\mathcal{U}} = \{ {U_i}|{U_i} \in F_p^*, i \in {I_n}\} $, ${I_n} = \{ 1,2,...,n\} $ and a dealer in the scheme. $\mathbf{LC}$ is a $[n + 1,t]$ linear code of length ($n+1$), dimension $t$. The dealer chooses two large primes $p$,$q$ with $p > n{q^2}$ and ${G_{t \times (n + 1)}} = ({\vec g_0},{\vec g_1},...,{\vec g_n})$ which is the public generator matrix of $\mathbf{LC}$. ${\vec g_i}$ is a column vector for $i = 0,1,...n.$ ${G_{t \times (n + 1)}}$ has rank $t$, i.e., any $t$ column vectors are linearly independent while any set of $t+1$ column vectors are linearly dependent, which guarantees that any $t$ or more shareholders are qualified to reconstruct the secret, but less than $t$ shareholders are unqualified. The following Vandermonde matrix is an option of ${G_{t \times (n + 1)}}$ for distinct ${U_i} \in F_p^*,i = 0,1,...,n$. 
\begin{equation} \label{Equation2}
{G_{t \times (n + 1)}} = {\begin{bmatrix}
{{1}}& {1} & \ldots &{{1}}\\
{U_0} & {U_1} & \ldots &{{U_{n}}}\\
 \vdots & \vdots& \ddots & \vdots \\
{{(U_0)^{t-1}}}& {(U_1)^{t-1}} & \cdots &{{(U_{n})^{t-1}}}
\end{bmatrix}}
\end{equation}

The dealer firstly randomly chooses the secret $s \in {F_q}$ and determines a non-zero row vector $\vec v = ({v_0},...,{v_{t - 1}}) \in F_p^t$ privately such that $s = \vec v{\vec g_0}\bmod p$, and then generates the code word $\vec w = (s,{s_1},...,{s_n}) = \vec v{G_{t \times (n + 1)}}$. Finally, the dealer allocates ${s_i}$ to ${U_i}$ as the share secretly for $i = 0,1,...n.$ \\
\textbf{2)	Component Construction} \par
If $m$ $(m \ge t)$ shareholders, ${{\mathcal{U}}_m} = \{ {U_i}|i \in {I_m}\} $, $({I_m} \subseteq {I_n},|{I_m}| = m \ge t)$, need to recover the secret $s$, they determine the corresponding public coefficients $\{ {b_i}|{b_i} \in F_p^*,i \in {I_m}\} $ non-interactively such that ${\vec g_0} = \sum\limits_{i \in {I_m}} {{b_i}{{\vec g}_i}} \bmod p$ holds. $\{ {b_i}|{b_i} \in F_p^*,i \in {I_m}\} $ is easy to find because $\{ {\vec g_i}|i \in {I_m}\} $ and ${\vec g_0}$ are linearly dependent. Take ${I_m} = \{ 1,2,...,m\} $ for example. Each participant ${U_j} \in {{\mathcal{U}}_m}$ can independently determine $\{ {b_i}|{b_i} \in F_p^*,i \in {I_m}\} $ as follows. ${U_j}$ firstly let ${b_i} = 1$ for $i =1,...m-t$, and then evaluates the remaining $t$ coefficients ${b_{m-t+1},b_{m-t+2},...,b_m}$, such that ${\vec g_0} = (\sum\nolimits_{i = 1}^{m - t} {{\vec g}_i} + \sum\nolimits_{i = m-t+1}^{m } {{b_i}{{\vec g}_i})\bmod p} $. In this way, all participants share $\{ {b_i}|{b_i} \in F_p^*,i \in {I_m}\} $ without interaction. \par
Each shareholder ${U_i}$ picks a random number $ {r_i}{ \in _U}{F_q} $ in private and constructs a component $c_i$ matched with new threshold $m$ as ${c_i} = ({b_i}{s_i} + {r_i}q)\bmod p.$
\begin{remark}
As a matter of fact,  ${b_i}$ can be directly expressed as $\prod\limits_{j \in I_m-\{i\}} {\frac{{{U_0} - {U_j}}}{{{U_i} - {U_j}}}} \bmod p$ due to Lagrange interpolation if the dealer chooses ${G_{t \times (n + 1)}}$ as in Equation \ref{Equation2}. ${c_i} = ({b_i}{s_i} + {r_i}q)\bmod p$ means the component ${c_i}$ binds all the shareholders ${{\mathcal{U}}_m}$ together. Therefore, the threshold is changed from $t$ to $m$.
\end{remark}
\noindent \textbf{3)	Secret Reconstruction} \par
Each shareholder in ${{\mathcal{U}}_m}$, e.g. ${U_i}$, releases the component ${c_i}$ to the others. After obtaining all components $\{ {c_j}_{}|j \in {I_m}\} ,$ ${U_i}$ recovers the secret as $s = (\sum\nolimits_{j \in {I_m}} {{c_j}} \bmod p)\bmod q.$
\begin{figure*}
\begin{tabular}{p{16cm}}
  \hline
  \textbf{Entities:}\\
\quad Dealer: $D$;\\
\quad $n$ shareholders: $ {\mathcal{U}} = \{ {U_i}|{U_i} \in F_p^*,i \in {I_n}\} ,$ $ {I_n} = \{ 1,2,...,n\} ;$ ${U_i}$ is the public identity of each shareholder;\\
\quad $m$ out of $n$ shareholders: ${{\mathcal{U}}_m} = \{ {U_i}|i \in {I_m}\}, {I_m} \subseteq {I_n},|{I_m}| = m \ge t$; \\
  \textbf{Parameters:}\\
\quad Public primes: $p,q$ with \textbf{$p > n{q^2}$}; \\
\quad Linear code: $\mathbf{LC}$ with length $n+1$ and dimension $t$; \\
\quad Public generator matrix of $\mathbf{LC}$: ${G_{t \times (n + 1)}} = ({\vec g_0},{\vec g_1},...,{\vec g_n})$ with rank \emph{t}, column vector ${\vec g_i} \in F_p^t$, $i{\rm{ = }}0,1,...,n$; \\
\quad Private row vector: $\vec v = ({v_0},...,{v_{t - 1}}) \in F_p^t$; \\
\quad Secret:$s = \vec v{\vec g_0}\bmod p$, $s \in {F_q}$; \\
\textbf{Algorithms:}\\
\quad \emph{\textbf{1) Share Generation}} \\
\quad $D$ randomly picks $s \in {F_q}$, designates $\vec v = ({v_0},...,{v_{t - 1}}) \in F_p^t$ and ${G_{t \times (n + 1)}}$, such that  $s = \vec v{\vec g_0}\bmod p$, allocates ${s_i} = \vec v{\vec g_i}\bmod p$ to $ {U_i} $  as the share privately and securely for $i = 1,2,...,n$, makes ${G_{t \times (n + 1)}}$ public and keeps $\vec v$ and \emph{s} in secret.\\
\quad \emph{\textbf{2) Component Construction}} \\
\quad To recover the secret, $m$ shareholders should raise threshold from $t$ to $m$. That is, according to some specified rule (see step $\textit{2)}$ in Section \ref{scheme}), each participant ${U_i} \in {{\mathcal{U}}_m}$ first determines the unique set of public coefficients $\{ {b_i}|{b_i} \in F_p^*,i \in {I_m}\} $ by itself such that ${\vec g_0} = \sum\nolimits_{i \in {I_m}} {{b_i}{{\vec g}_i}} \bmod p$. Then, it picks a random number ${r_{{i_{}}}}{ \in _U}{F_q}$ privately to compute a component as ${c_i} = ({b_i}{s_i} + {r_{{i_{}}}}q)\bmod p$. \\
\quad \emph{\textbf{3) Secret Reconstruction}} \\
\quad Each shareholder ${U_i}_{} \in {{\mathcal{U}}_m}$ releases ${c_i}$ to the others. After collecting all components, each shareholder independently recovers the secret as $s = (\sum\nolimits_{j \in {I_m}} {{c_j}} \bmod p)\bmod q$. \\
  \hline
\end{tabular}
\caption{The proposed TCSS scheme} \label{Fig2}
\end{figure*}

\subsection{Correctness}
\begin{theorem} \label{theorem1}
In the proposed TCSS scheme, any $t$ or more than $t$ shareholders are able to reconstruct the secret from all their components. That is, given $m$ $(m \ge t)$ shareholders ${{\mathcal{U}}_m} = \{ {U_j}|j \in {I_m}\} $, each shareholder ${U_j}$ with the component ${c_j} = ({b_j}{s_j} + {r_j}q)\bmod p,j \in {I_m}$, the secret can be recovered as $s = (\sum\nolimits_{j \in {I_m}} {{c_j}} \bmod p)\bmod q.$
\end{theorem}
\textbf{proof} \\\\

\begin{align}
& (\sum\nolimits_{j \in {I_m}} {{c_j}} \bmod p)\bmod q {}\nonumber \\
 &= (\sum\nolimits_{j \in {I_m}} {{b_j}{s_j}} \bmod p + \sum\nolimits_{j \in {I_m}} {{r_j}q} )\bmod p\bmod q  {}\nonumber \\
 &= (s + \sum\nolimits_{j \in {I_m}} {{r_j}q} )\bmod p\bmod q \label{equation3} \\
 &= (s + \sum\nolimits_{j \in {I_m}} {{r_j}q} )\bmod q  \label{equation4} \\
 &= s  {}\nonumber 
\end{align}

Note that we have $s + \sum\nolimits_{j \in {I_m}} {{r_j}q}  \le s + m(q - 1)q < n{q^2} < p$ due to $s \in {F_q}$, ${r_j}{ \in _U}{F_q}$ and $p > n{q^2}.$ As a result, Equation \ref{equation3} is equivalent to Equation \ref{equation4}.
\quad \\
\section{Security Analyses and Performance Comparison}
\subsection{Security analyses}
In the TCSS scheme, components instead of original shares are used to raise threshold. In the following, we give Lemma \ref{l1}, Lemma \ref{l2} and Corollary \ref{c1} as the basis. Then,  We show the security by three theorems. Theorem \ref{thr2} testifies to the fact that up to $t-1$ shareholders are still unable to reconstruct the secret. Theorem \ref{thr3} guarantees the scheme is resistant to IPA. Theorem \ref{thr4} proves the threshold in our proposed scheme is changeable.
\begin{lemma} \label{l1}
Suppose that random variable $x$ is uniformly distributed in ${F_p}$, for any value $t \in F_p^*$. $xt$ has a uniform distribution over ${F_p}$.
\end{lemma}
\textbf{proof} \\\\
We immediately get the lemma from the property of finite field ${F_p}$. 
\quad \\

\begin{lemma} \label{l2}
Given prime $p$ and random variables ${x_i}$ mutually independent and uniformly distributed in ${F_p}$, $\sum\nolimits_{i = 1}^k {{t_i}{x_i}} \bmod p$ has a uniform distribution over ${F_p}$, if not all values ${t_i} \in {F_p}$ are zero, where $i = 1,2,...,k$.
\end{lemma}
\textbf{proof} \\\\
Let us first consider the case of $k = 2$ then generalize the case of $k$ being any positive integer.
\begin{enumerate}
\item If ${t_1}$ or ${t_2}$ is zero, it is obvious that $({t_1}{x_1} + {t_2}{x_2})\bmod p$ is uniformly distributed over ${F_p}$ from Lemma \ref{l1}. \
\item If both ${t_1}$ and ${t_2}$ are nonzero, ${t_1}{x_1}$ and ${t_2}{x_2}$ are uniformly distributed over ${F_p}$ from Lemma \ref{l1}. To prove $({t_1}{x_1} + {t_2}{x_2})\bmod p$ is uniformly distributed in ${F_p}$, we assume that ${x_{11}}$ and ${x_{12}}$ are any two different values of variable ${x_1}$. In this case, ${t_1}{x_{11}}$ and ${t_1}{x_{12}}$ are obviously distinct in ${F_p}$ for $\gcd ({t_1},p) = 1$. Thus, $({t_1}{x_{11}} + {t_2}{x_2})\bmod p$ and $({t_1}{x_{12}} + {t_2}{x_2})\bmod p$ are two distinct permutations of $\{ 0,1,2,...,p - 1\} $ when ${x_2}$ varies over ${F_p}$. More generally, $({t_1}{x_{1i}} + {t_2}{x_2})\bmod p$ are distinct permutations of $\{ 0,1,2,...,p - 1\} $ for different ${x_{1i}}$ which values of random variable ${x_1}$. That is, each value in $\{ 0,1,2,...,p - 1\} $ appears with the same frequency. Therefore, ${t_1}{x_1} + {t_2}{x_2}$ is uniformly distributed in ${F_p}$.
\end{enumerate}
Now, $({t_1}{x_1} + {t_2}{x_2})\bmod p$ is uniformly distributed in ${F_p}$. By iterating the procedure, we have that $\sum\nolimits_{i = 1}^k {{t_i}{x_i}} \bmod p$ has a uniform distribution over ${F_p}$. 
\quad \\

\begin{corollary}\label{c1}
$(\sum\nolimits_{i = 1}^k {{a_i}{x_i}}  + \sum\nolimits_{j = 1}^l {{b_j}{y_j}} )\bmod p$ has a uniform distribution over ${F_p}$, if random variables ${x_i}$ and ${y_i}$ are uniformly distributed over ${F_p}$ and ${F_q}$ respectively for $i = 1,2,..,k$ and $ j = 1,2,..,l$, where all variables are mutually independent. Besides, $p$ and $q$ are positive primes with $q \le p$ and ${a_i},{b_j} \in F_p^*.$ 
\end{corollary}
\textbf{proof} \\\\
(Omitted) The corollary can be proved by the method similar to Lemma \ref{l2}.  \\
\quad \\

\begin{theorem} \label{thr2}
In the TCSS scheme, less than $t$ shareholders cannot obtain the secret. That is, less than $t$ shareholders obtain nearly no information about the secret $s$, i.e. ${\lim _{q \to  + \infty }}I(s;{{\Omega}}') = 0, $ where ${{\Omega'}}$ denotes a set of less than $t$ shares. 
\end{theorem}
\textbf{proof} \\\\
For simplicity, assume that ${{\mathcal{U}}_{t - 1}} = \{ {U_1},{U_2},...,{U_{t - 1}}\} $ are the $(t-1)$ shareholders with the corresponding shares $\Omega_{t-1}=\{{s_1},{s_2},...,{s_{t - 1}}\}$, while ${s_t} = \vec v{\vec g_t}\bmod p$ is the unknown share for ${{\mathcal{U}}_{t - 1}}$ and the secret can be computed as $s = \sum\nolimits_{i = 1}^t {{b_i}{s_i}} \bmod p\bmod q$, where ${b_i}$ can be publicly determined from the generator matrix. We assume $\mathcal F(.)$ is any function which takes $\Omega_{t-1}$ or its subset as input and produces a presumed secret $s'$ as output. Suppose that ${{\mathcal{U}}_{t - 1}}$ derives a presumed secret  $s'=\mathcal{F}(\Omega_{t-1})$. Let us examine the probability $P(s|\Omega_{t-1})$, i.e., $P(s'=s)$.
\begin{align}
s'= s =\sum\nolimits_{i = 1}^t {{b_i}{s_i}} \bmod p \bmod q  {} \nonumber\\
 \to \sum\nolimits_{i = 1}^t {{b_i}{s_i}} \bmod p -s' =\lambda q  \label{eqt-1}
\end{align}

From Lemma \ref{l2}, the left side of Equation \ref{eqt-1} is uniformly distributed over ${F_p}$ for ${{\mathcal{U}}_{t - 1}}$ since the unknown ${s_t}$ has a uniform distribution over ${F_p}$. Similarly, for any less than $t$ shareholders with the corresponding share set ${\Omega'}$, the property of uniform distribution over ${F_p}$ still holds. Consequently, there are at most $\left\lfloor {p/q} \right\rfloor  + 1$ values of $\lambda $ satisfying Equation \ref{eqt-1} in ${F_p}$. It means less than $t$ shareholders obtain the secret with the largest probability $(\left\lfloor {p/q} \right\rfloor  + 1)/p$, namely, $P(s|\Omega') = (\left\lfloor {p/q} \right\rfloor  + 1)/p$. As a result, for any small value $\varepsilon$, we have
\begin{align*}
I(s;\Omega') & = H(s) - H(s|\Omega') \le \log q - \log \frac{p}{{\left\lfloor {p/q} \right\rfloor  + 1}} \\
& = \log \frac{{q(\left\lfloor {p/q} \right\rfloor  + 1)}}{p} < \log \frac{{p + q}}{p} \\ 
& = \log \frac{{p/q + 1}}{{p/q}} < \varepsilon.
\end{align*}

Thus, ${\lim _{p/q \to  + \infty }} I(s;\Omega') = 0,$ i.e., ${\lim _{q \to  + \infty }}I(s;\Omega') = 0$ due to $p/q > nq$. 
\quad \\

\begin{theorem} \label{thr3}
The proposed TCSS scheme is able to thwart IPA. Concretely, if an adversary without valid share participate in secret reconstruction with no less than $t$ legal shareholders, the secret cannot be recovered. 
\end{theorem}
\textbf{proof} \\\\
We first consider the normal case that $m$ participants, e.g. ${{\mathcal{U}}_m} = \{ {U_1},{U_2},...,{U_m}\} $ for simplicity, construct the corresponding components $\{ {c_1},{c_2},...,{c_m}\}$ to recover the secret. That is, each participant ${U_i} \in {{\mathcal{U}}_m}$ constructs the component ${c_i} = {b_i}{s_i} + {r_i}q = ({b_i}\vec v{\vec g_i} + {r_i}q)\bmod p$ with the share ${s_i}$. Thus, $({c_1},{c_2},...,{c_m}) = \vec v({b_1}{\vec g_1},{b_2}{\vec g_2},...,{b_m}{\vec g_m}) + ({r_1}q,{r_2}q,...,{r_m}q)\bmod p$, where each ${b_i}$ is public while $\vec v$ is the secret and ${r_i}$ is private. \par
Suppose an adversary impersonates the shareholder ${U_m}$, i.e., it does not have  $s_m$ which is the valid share of $U_m$, but can communicate with the other legal shareholders. In the following, we first show that, $c_m$, the component of $U_m$, is indistinguishable from a random variable uniformly distributed in $F_p$ for the adversary. Then, we prove that using a forged component to recover the secret is nearly as difficult as directly guessing the secret within the secret space. 
\begin{enumerate}
\item We have ${s_i} = \vec v{\vec g_i}$ with non-zero vector $\vec v = ({v_0},...,{v_{t - 1}}) \in F_p^t$ and ${\vec g_i} = (1,{U_i},U_i^2,...,U_i^{t - 1})^{\rm T} \in F_p^t$, where each ${v_i}$ is an integer uniformly and privately selected by the dealer within ${F_p}$. From the view of the adversary, each ${v_i}$ is indistinguishable from a random variable uniformly distributed over ${F_p}$. ${\vec g_i}$ is a nonzero column vector for ${U_i} \in F_p^*$. According to Lemma \ref{l2}, each unknown share, ${s_i} = \vec v{\vec g_i}\bmod p$, is uniformly distributed over ${F_p}$ for the adversary. Obviously, the corresponding component ${c_i} = ({b_i}{s_i} + {r_i}q)\bmod p$, and $\sum\nolimits_{i = 1}^m {{c_i}} \bmod p$ are also uniformly distributed over ${F_p}$ according to Corollary \ref{c1}. \
\item Because $c_m$ is indistinguishable from a random variable uniformly distributed in $F_p$ for the adversary, we write the forged component which the adversary uses to participate in secret reconstruction as $c_m'$ and $c_m'$ can be any value in $F_p$. Assume $\mathcal F(.)$ is any function which takes $\mathcal{C}_m'={\{c_1,c_2,...,c_{m-1}, c_m'\}}$ or any subset of $\mathcal{C}_m'$ as input and produces a presumed secret $s'=\mathcal{F}(\mathcal{C}_m')$ as output.  Without losing the generality, suppose that a participant derives a value $s'=\mathcal F(\mathcal{C}_m')$ in some way. If $s'$ happens to equal the secret $s$, the secret is recovered. Now let us examine the probability of $P(s'=s)$. 
\begin{align}
& s'=s=  (\sum\nolimits_{i = 1}^{m - 1} {{c_m}}  + c_m)\bmod p\bmod q \nonumber \\
& \to s'-(\sum\nolimits_{i = 1}^{m - 1} {{c_m}}  + c_m')\bmod p = \lambda q, \lambda\in Z  \label{eqth3}
\end{align}

Note that the left side of Equation \ref{eqth3} is uniformly distributed over $F_p$. Consequently, there are at most $\left\lfloor {p/q} \right\rfloor  + 1$ possible values of $\lambda $ satisfying Equation \ref{eqth3}. Hence, $P(s'=s)$ is not larger than $(\left\lfloor {p/q} \right\rfloor  + 1)/p$. Due to ${\lim _{q \to  + \infty }}(\left\lfloor {p/q} \right\rfloor  + 1)/p = 1/q$, using a forged component to recover the secret is nearly as difficult as directly guessing the secret within ${F_q}$. 
\end{enumerate}

As a result, the proposed TCSS scheme is able to thwart IPA. 
\quad \\

\begin{table*}
\caption{The performance comparisons} \label{tab1}
\begin{tabular}{l l l l l l l}
\hline
Scheme & Share size & Message size in secret reconstruction & Security & Recovery complexity \\ \hline
scheme \cite{martin1999changing} & $kH(s)$ & $H(s)$ & perfect & $O(m {\log ^2} t)$ \\
scheme \cite{blundo1996fully} & $H(s)$ & $(n-m+1)H(s)$ & perfect  & $O(n {\log ^2} n)$ \\
scheme \cite{harn2015dynamic} & $tH(s)$ & $2 \log k ^*$ & broken  & $O(m {\log ^2} m)$ \\
scheme \cite{zhang2012threshold} & $H(s)$ & $H(s)$ & computational & $O(m {\log ^2} m)$ \\
scheme \cite{steinfeld2006lattice} \cite{steinfeld2007lattice} & $H(s)$ & $H(s)$ & probabilistic & $O(2^{(1+\frac{1}{\epsilon})(k+t)})^{**}$ \\
scheme \cite{jia2019new} & $c(H(s)+k)^{***}$ & $c(H(s)+k)$ & computational & $O(m)$ \\
Our scheme & $2H(s)$ to $3H(s)$ & $2H(s)$ to $3H(s)$ & asymptotically perfect & $O(m {\log ^2} m)$ \\ \hline
\end{tabular}
\begin{tablenotes}\scriptsize
\item $t$: original threshold; $n$: the number of all the shareholders; $m$: the raised threshold value\
\item $^*:$ The greatest threshold is $k$.\
\item $^{**}:$ $\epsilon= \gamma_{{CVP}} - 1$, where $\gamma_{{CVP}}$ is the approximation factor of CVP approximation algorithm. \
\item $^{***}:$ $c = 1 + \frac{k-m+t}{t}$.
\end{tablenotes}
\end{table*}

\begin{theorem} \label{thr4}
The threshold in our proposed scheme is changeable. Concretely, if $m$ ($m > t$) shareholders participate in secret reconstruction, the secret must be recovered from all the $m$ components $\mathcal{C}_m=\{ {c_1},{c_2},...,{c_m}\} $, while nearly no information about the secret can be derived from no more than $m-1$ components, i.e., \[{\lim _{q \to  + \infty }}I(s;\mathcal{C}_j) = 0 \quad for \quad {\mathcal{C}_j} \subset {\mathcal{C}_m}.\]
\end{theorem}
\textbf{proof} \\\\
Define $\mathcal{C}_j$ as ${\mathcal{C}_j} = \{ {c_1},{c_2},...,{c_i},...,{c_j}\} $ with ${c_i} = ({b_i}{s_i} + {r_i}q)\bmod p$, ${r_i}{ \in _U}{F_q}$, $j < m$. Suppose a shareholder derives a presumed secret $s'=\mathcal F(\mathcal{C}_j)$, where $\mathcal F(.)$ is any function which takes $\mathcal{C}_j$ or its subset as input and produces a presumed value $s'$ as output. Successfully recovering the secret from ${\mathcal{C}_j}$ means $s'=s=\sum\nolimits_{i = 1}^m {{c_i}} \bmod p\bmod q$, namely, 
\begin{equation} \label{eqth4}
(\sum\nolimits_{i = 1}^m {{c_i}} \bmod p- s')\bmod p = \lambda q,\lambda  \in F_p. 
\end{equation}
Note that the left side of Equation \ref{eqth4} is uniformly distributed over ${F_p}$ according to Lemma \ref{l2}. Consequently, there are at most $\left\lfloor {p/q} \right\rfloor  + 1$ valid values of $\lambda $ satisfying Equation \ref{eqth4}. As a result, ${P(s|C_J)}$ is no larger than $(\left\lfloor {p/q} \right\rfloor  + 1)/p$. Note that the secret $s$ is uniformly selected from ${F_q}$, i.e., the probability $P(s) = 1/q$. Then, for any small value $\varepsilon$, the mutual information of $s$ with ${\mathcal{C}_j}$ is 
\begin{align*}
I(s;\mathcal{C}_j) & = H(s) - H(s|\mathcal{C}_j) \le \log q - \log \frac{p}{{\left\lfloor {p/q} \right\rfloor  + 1}} \\ 
& = \log \frac{{q(\left\lfloor {p/q} \right\rfloor  + 1)}}{p} < \log \frac{{p + q}}{p} \\
& = \log \frac{{p/q + 1}}{{p/q}} < \varepsilon. 
\end{align*}

Thus, ${\lim _{p/q \to  + \infty }} I(s;\mathcal{C}_j) = 0,$ i.e., ${\lim _{q \to  + \infty }}I(s;\mathcal{C}_j) = 0$ due to $p/q > nq$. 
\quad \\

\subsection{Performance Comparison}
In this subsection, we compare our proposed scheme with some other TCSS schemes \cite{blundo1996fully} \cite{harn2015dynamic} \cite{jia2019new} \cite{martin1999changing} \cite{steinfeld2006lattice} \cite{steinfeld2007lattice} and \cite{zhang2012threshold} in four aspects: share size, message size in secret reconstruction, security and recovery complexity. The results are shown in Table \ref{tab1}.\par
From Table \ref{tab1}, we can see that schemes \cite{blundo1996fully} \cite{steinfeld2006lattice} \cite{steinfeld2007lattice} \cite{zhang2012threshold} have ideal share size. In our scheme, the secret $s$ is in $F_q$ while the share $s_i$ is in $F_p$, where $q^3 > p > nq^2$ and $q$ is much greater than $n$. Therefore, the share size is between $2H(s)$ and $3H(s)$. Message size in secret reconstruction means size of the message which a shareholder sends to others. In our scheme, it is totally equal to the share size since shareholders send components as the message which is also in $F_p$. In terms of security, schemes \cite{blundo1996fully} and \cite{martin1999changing} can be perfect. Schemes \cite{jia2019new} and \cite{zhang2012threshold} are computational security since they need encryption algorithms. The security of schemes \cite{steinfeld2006lattice} and \cite{steinfeld2007lattice} is based on CVP problem, and thus the two scheme are probabilistic security. Besides, as we mentioned in Section 1, scheme \cite{harn2015dynamic} has been attacked successfully by paper \cite{jamshidpour2017security}. Our proposed TCSS scheme is independent of one way function, encryption or conventional hard problems and Theorem \ref{thr2} \ref{thr3} \ref{thr4} show it is asymptotically perfect security. Finally, because our scheme is based on linear code, the recovery complexity  is also $O(m {\log ^2} m)$ like most linear SS scheme.
\section{Application to Group Authentication}
\begin{figure*}
\begin{tabular}{p{16cm}}
  \hline
  \textbf{Entities:}\\
\quad The Group Manager: GM;\\
\quad Complete set of $n$ group members:${\mathcal{U}} = \{ {U_i}|{U_i} \in F_p^*,i \in {I_n}\} {,_{}}{I_n} = \{ 1,2,...,n\} ;$ ${U_i}$ is the public identity of each group member.\\
  \textbf{Parameters:}\\
\quad Public primes: $p,q$ with \textbf{$p > n{q^2}$}; \\
\quad Linear code: $\mathbf{LC}$ with length $n+1$ and dimension $t$; \\
\quad Public generator matrix of $\mathbf{LC}$: \\
\quad ${G_{t \times (n + 1)}} = ({\vec g_0},{\vec g_1},...,{\vec g_n})$ with rank \emph{t}, column vector ${\vec g_i} \in F_p^t$, $i{\rm{ = }}0,1,...,n$; \\
\quad Private row vector: $\vec v = ({v_0},...,{v_{t - 1}}) \in F_p^t$; \\
\quad Secret:$s = \vec v{\vec g_0}\bmod p$, $s \in {F_q}$; \\
\quad One way hash function: $H(\cdot)$; \\
\textbf{Algorithms:}\\
\quad \emph{\textbf{1)	Token Generation}} \\
\quad GM designates $\vec v$, $s$ and ${G_{t \times (n + 1)}}$, allocates ${s_i} = \vec v{\vec g_i}$ to ${U_i}$, $i = 1,2,...n$, as the token securely, makes $H(s)$, ${G_{t \times (n + 1)}}$ public while keeps $\vec v$ and $s$ secret.\\
\quad \emph{\textbf{2)	Group Authentication}} \\
\quad (1) To authenticate any $m$ users ${{\mathcal{P}}_m} = \{ {P_i}|i \in {I_m} \subseteq {I_n},|{I_m}| = m \ge t\} $ at once, each user ${P_i} \in {{\mathcal{P}}_m}$ determines the public coefficients $\{ {b_i}|{b_i} \in F_p^*,i \in {I_m}\} $ independently such that ${\vec g_0} = \sum\nolimits_{i \in {{\rm{I}}_m}} {{b_i}{{\vec g}_i}} \bmod p$, and then picks random number ${r_{{i_{}}}}{ \in _U}{F_q}$ privately to compute a component as ${c_i} = ({b_i}{s_i} + {r_{{i_{}}}}q)\bmod p$.\\
\quad (2) Each user ${P_i} \in {{\mathcal{P}}_m}$ releases the component ${c_i}$ to the other users through private channels. After collecting all components, each user computes $s'$ as $s' = (\sum\nolimits_{j \in {I_m}} {{c_j}} \bmod p)\bmod q$. If $H(s') = H(s)$, all users in ${{\mathcal{P}}_m}$ have been authenticated successfully, i.e., all users in ${{\mathcal{P}}_m}$ belong to ${\mathcal{U}}$. Otherwise, there is at least one non-member of $\mathcal{U}$. \\
  \hline
\end{tabular}
\caption{Group authentication scheme based on TCSS} \label{Fig3}
\end{figure*}

To facilitate authentication in group oriented applications such as group chat in Wechat, Harn proposed the notion of group authentication \cite{harn2012group}, which allows each group user to check whether all users belong to the same group at once. Concretely, group authentication can be formulated as follows. The group manager GM computes tokens $s_i$ from a selected secret $s$, allocates each token $s_i$ to group member $U_i$ securely and makes $H(s)$ publicly known, where $H(\cdot)$ is a one-way hash function.  In a group authentication scheme, there are $m$ users $P_i, i=1,2,...,m$. Each user $P_i$ computes a component $c_i$ from its token and releases  $c_i$  to the others.  Group authentication allows each user to verify whether all released values are valid at once. That is, 
\begin{align*}
GA\{ H(s) & \stackrel{?}{=} H(F(c_1,c_2,...,c_m))\}  \\
& = \left\{ \begin{array}{l}0 \rightarrow \exists P_i \notin U, i=1,2,...,n;\\1 \rightarrow \forall P_i \in U, i=1,2,...,n,\end{array} \right.
\end{align*}

where $GA$ is the group authentication algorithm and $F$ is a public function.\par
In Harn's scheme \cite{harn2012group}, GM selects $k$ polynomials and requires $kt>n-1$ to guarantee the security, where $t$ is the least number of participants and $n$ is the number of all the users. However, \cite{ahmadian2017linear} has proved that $m$ components are linearly dependent in the case of $k+t-1 < m < kt$. Therefore, Harn's scheme is insecure. Based on the notion of group authentication, Mahalle et al. \cite{mahalle2014threshold} developed a group authentication scheme, called TCGA, for the Internet of Things (IoT), which uses Paillier threshold cryptography as the underlying secret sharing scheme. Paillier threshold cryptography is a public key variant of the $(t,n)$ threshold SS. However, it is doubtful whether or not the TCGA scheme is appropriate for IoT because most nodes in IoT are low in computing power.\par
In the following, we will propose a group authentication scheme based on the above TCSS scheme, which allows each group user to independently check whether all users belong to the same group at once. 
\subsection{Group authentication scheme based on TCSS}
In our TCSS scheme, when each group member is allocated a share as the token in advance and all users (group members or non-members) collectively run the secret reconstruction algorithm, the secret can be recovered only if each user has a valid token. That is, recovering the secret successfully means all users are legal group members. In this way, any user is able to authenticate that whether all users are legal members at a time, instead of one by one.\par
In Figure \ref{Fig3}, we present a new group authentication scheme base on the proposed TCSS scheme. It consists of two algorithms, token generation and group authentication.
\begin{remark}
Based on TCSS, the group authentication scheme can be easily converted into an authenticated group key agreement scheme by modifying step (2) in group authentication as follows. \par
Each user ${P_i} \in {{\mathcal{P}}_m}$ releases the component to the other users through private channels. On collecting all components, each user computes $s' = (\sum\nolimits_{j \in {I_m}} {{c_j}}  \bmod p)\bmod q$. If $H(s') = H(s)$, all users in ${{\mathcal{P}}_m}$ belong to ${\mathcal{U}}$ and the group key is $k = \sum\nolimits_{j \in {I_m}} {{c_j}} \bmod p$. Otherwise, there exists at least one non-member of ${\mathcal{U}}$ and the group key agreement is aborted. \par
\end{remark}
\subsection{Correctness and security analyses}
In the proposed group authentication, a user can recover the secret and thus authenticate all users at once if each user is a group member and releases a valid component. Otherwise, if a user (i.e. non-member) does not have a valid token, it cannot construct a valid component. Consequently, the secret cannot be recovered correctly and the non-member fails to pass the group authentication. 
\begin{theorem}
In the proposed group authentication scheme, a non-member, even with $m-1$ valid components of a group, fails to forge a new valid component of the group to pass the authentication. 
\end{theorem}
\textbf{proof} \\\\
If a non-member, with $m-1$ valid components, could forge a valid component, it must be capable of recovering the secret. However, it follows from Theorem \ref{thr3} that a non-member is unable to obtain the secret. Therefore, the non-member fails to forge a valid component.
\quad \\

\begin{theorem}
In the proposed group authentication scheme, $t-1$ group members fails to forge a valid token for a non-member to pass the authentication in the same group.
\end{theorem}
\textbf{proof} \\\\
The theorem can be immediately obtained from Theorem \ref{thr2}. 
\quad \\
\subsection{Properties}
Obviously, the communication overhead is limited because each user merely needs to release a component to the others. This can be efficiently accomplished by broadcasting the component if all users share a private broadcasting channel. In authentication, each user only needs to compute a component from the token locally and releases it to the other users, and then adds its own component and $m-1$ received ones up to authenticate all users. Besides, the proposed group authentication scheme does not depend on any public key system and thus is more efficient in computation compared with TCGA \cite{mahalle2014threshold}. Moreover, distinct from conventional authentication schemes which authenticate a single user each time, our group authentication scheme authenticates all users at once. \par
As a matter of fact, group authentication scheme can be constructed simply based on traditional $(t,n)$ threshold SS if all users (i.e., participants in $(t,n)$ threshold SS) release their tokens (i.e., shares) at the same time. In this case, a non-member will be detected since it releases a wrong token. In contrast, the proposed group authentication works for $m$ ($t \le m \le n$) users and does not require all users to release components simultaneously. Moreover, the authentication scheme allows each user to hold only one token and does not have limitation, such as $kt>n-1$ in Harn's group authentication scheme.
\begin{remark}
Of course, the proposed group authentication scheme can only verify whether all users are group members, but cannot identify which one is non-number. However, the scheme can be used to pre-authenticate all users efficiently. Then, conventional authentication schemes can be applied if there is a non-member.
\end{remark} 
\section{Conclusions}
Nowadays, group oriented applications are getting more and more popular and require more secure and efficient SS scheme to satisfy their security requirements. The paper utilizes linear code to construct a TCSS scheme. In the proposed scheme, shareholders can change the initial threshold from $t$ to $m$, where $m$ is the exact number of shareholders who participate in secret reconstruction. In this way, the secret can be recovered only if all the participants are legal shareholders, which means our scheme is resistant to IPA. Besides, the TCSS scheme is independent of one way function, encryption or conventional hard problems and realizes asymptotically perfect security. To complete rapid m-to-m authentication in group oriented applications, a group authentication scheme is proposed based on the TCSS scheme. It allows each user in a group to check whether all users are legal group members at once.

\bibliographystyle{plain}

\bibliography{cas-refs}

\end{document}